\documentclass[%
reprint,
superscriptaddress,
amsmath,amssymb,
aps,
prb,
]{revtex4-1}
\setcitestyle{numbers,square}

\usepackage{braket}
\usepackage{float}
\usepackage{placeins}
\usepackage{graphicx}
\usepackage{dcolumn}
\usepackage{bm}
\usepackage[normalem]{ulem} 
\usepackage{xcolor}

\usepackage{hyperref}
\hypersetup{
    colorlinks,%
    citecolor=blue,%
    linkcolor=blue,%
    urlcolor=blue
}

\newcommand{\orcid}[1]{\href{https://orcid.org/#1}{\includegraphics[width=8pt]{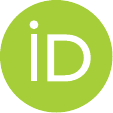}}}

\begin{document}

\title{Design of spin-orbital-textures in ferromagnetic/topological insulator interfaces}

\author{A. L. Araújo\orcid{0000-0002-6835-6113}}
\affiliation{Ilum School of Science, Brazilian Center for Research in Energy and Materials (CNPEM), Campinas, SP, Brazil}

\author{F. Crasto de Lima\orcid{0000-0002-2937-2620}} 
\email{felipe.lima@ilum.cnpem.br}
\affiliation{Ilum School of Science, Brazilian Center for Research in Energy and Materials (CNPEM), Campinas, SP, Brazil}

\author{C. H. Lewenkopf\orcid{0000-0002-2053-2798}} 
\affiliation{Instituto de Física, Universidade Federal Fluminense, Niterói, RJ, Brazil}

\author{A. Fazzio\orcid{0000-0001-5384-7676}}
\email{adalberto.fazzio@ilum.cnpem.br}
\affiliation{Ilum School of Science, Brazilian Center for Research in Energy and Materials (CNPEM), Campinas, SP, Brazil}

\date{\today}

\begin{abstract}

Spin-orbital textures in topological insulators due to the spin locking with the electron momentum, play an important role in spintronic phenomena that arise from the interplay between charge and spin degrees of freedom. We have explored interfaces between a ferromagnetic system (CrI$_3$) and a topological insulator (Bi$_2$Se$_3$) that allow the manipulation of spin-orbital textures. Within an {\it ab initio} approach we have extracted the spin-orbital-textures dependence of experimentally achievable interface designs. The presence of the ferromagnetic system introduces anisotropic transport of the electronic spin and charge. From a parameterized Hamiltonian model we capture the anisotropic backscattering behavior, showing its extension to other ferromagnetic/topological insulator interfaces. We verified that the van der Waals TI/MI interface is an excellent platform for controlling the spin degree of freedom arising from topological states, providing a rich family of unconventional spin texture configurations.

\end{abstract}

\maketitle

\section{Introduction}
\label{sec:introduction}

Recent years have witnessed an enormous growth in the interest in the role played by symmetry, dimensionality, and topology in quantum properties of condensed matter and material science systems. The study and manipulation of the spin-orbit coupling (SOC) and the spin degree of freedom is key for advancing this line of research and revealed a plethora of new fascinating phenomena, such as quantum Hall effects \cite{RMPSinova2015}, charge/spin interplay effects \cite{REE, IREE}, that lead to many spintronic applications \cite{n1}.

Since SOC occurs through the response of the electron's spin to the nuclear electric field or in response to a polarization field resulting from symmetry breaking, efficient control of the spin configuration of electronic states has been achieved by manipulation of their inversion and/or time-reversal (TRS) symmetries. For instance, in materials without inversion symmetry, spin polarization is accompanied by a spin splitting of electronic states and the appearance of a chiral spin texture due to a spin-momentum locking, thought Rashba \cite{n9, n10} and Dresselhaus \cite{n11} effects.

The SOC also plays a central role in the topological states of matter, where the nontrivial phase of the bulk is usually due to a band inversion at the Fermi level. Topological insulators (TI) are characterized by the presence of insulating bulk states but metallic states at the interfaces with topologically trivial systems \cite{TI-1, TI-2}, which also exhibit a spin-momentum locking and a helical spin texture with opposite chiralities of spin. These states are robust against external perturbations due to the presence of TRS, which suppresses backscattering processes. 

One of the main characteristics of chiral spin textures is that they can be exploited to generate out-of-equilibrium spin polarizations, a process directly linked to one of the primary objectives of spintronics: at room temperature achieving an efficient mechanism for interconversion between pure spin excitation and an electrical signal \cite{Spintronic_1, Spintronic_2}, with the Rashba-Edelstein (REE) \cite{REE} and the inverse Rashba-Edelstein (IREE) \cite{IREE} effects being among the most studied. Additionally, another challenge faced by spintronics is to increase the carrier lifetime by suppressing the spin relaxation interaction. One path to achieve this goal is to obtain a material where the spin configuration is uniform in space and independent of the moment, configuring the persistent spin helix (PSH), where electron motion is accompanied by spin precession \cite{NATCOMMtao2018}. The advancement of spintronics is not only closely tied to the ability to obtain spin-textured surface states but primarily to the capacity to manipulate the configuration of such states. Substantial effort is currently dedicated to this task, particularly concerning the investigation of spin textures originating from topological materials (such as topological insulators \cite{TI-1, TI-2, TCI-1, TCI-2}, Weyl semimetals \cite{Weyl-1, Weyl-2}) or materials with inversion asymmetry. 

In this context, in recent years research on Topological Insulator/Magnetic Insulator (TI/MI) interfaces for spintronic applications has been driven by the quest of efficient charge-spin interconversion mechanisms \cite{n15} and spin torque transfer \cite{STT-1, STT-2}, the latter being a promising effect for the development of non-volatile and low-power magnetic memories \cite{n17}. Additionally, for practical applications, TI/MI heterostructures with van der Waals (vdW) coupling are excellent candidates for spintronics, as they preserve the topological properties and spin texture behavior of TI surface states, suppressing the occurrence of interfacial charges while protecting the material lattice avoiding the formation of structural defects \cite{n18}. 

In this paper, we propose using a Topological Insulator/Magnetic Insulator van der Waals interface to enable the creation and design of a large variety of topological spin-textures. Our proposal is based on a systematic study of the electronic states of a CrI$_3$/Bi$_2$Se$_3$ interface using realistic {\it ab initio} calculations. We choose the CrI$_3$/Bi$_2$Se$_3$ interface as a study platform due to its amenable synthesis and recent experimental interest \cite{CrI3-3, CrI3-4, bi2se3-mbe-1, bi2se3-mbe-2}. In addition, Bi$_2$Se$_3$ is one of the most theoretically and experimentally studied three-dimensional topological insulators \citep{Bi2Se3-TI-4, Bi2Se3-TI-5, Bi2Se3-TI-1, Bi2Se3-TI-2, Bi2Se3-TI-3}, whereas  CrI$_3$ stands as a robust ferromagnetic (FM) insulator \cite{CrI3-1, CrI3-2, CrI3-4}, being a promising platform for two-dimensional magnetism due to its magnetic anisotropy \cite{CrI3-1, CrI3-2}, and all-optical control of magnetization \cite{CrI3-6}. We provide an interpretation of the results employing a simple effective model that captures the main features of coupling between the TI and FM surfaces. Further, we discuss some features of the disorder scattering of electrons in TI states depending on the FM magnetization direction.

This paper is organized as follows: In Sec. \ref{sec:methods} we present the geometry of the CrI$_3$/Bi$_2$Se$_3$ interface, the methodology used in first-principles calculations, as well as the effective Hamiltonian model that describes the interface states and the behavior of the scattering processes. In Sec. \ref{sec:Results-discussion}, we analyze the behavior and properties of spin-orbital textures and scattering rates as a function of different interface degrees of freedom. We present our conclusions in Sec. \ref{sec:conclusions}.

\section{Methods}
\label{sec:methods}

\subsection{CrI$_3$/Bi$_2$Se$_3$ Interface}
\label{sec:interface}

\begin{figure}[t]
    \includegraphics[width=\columnwidth]{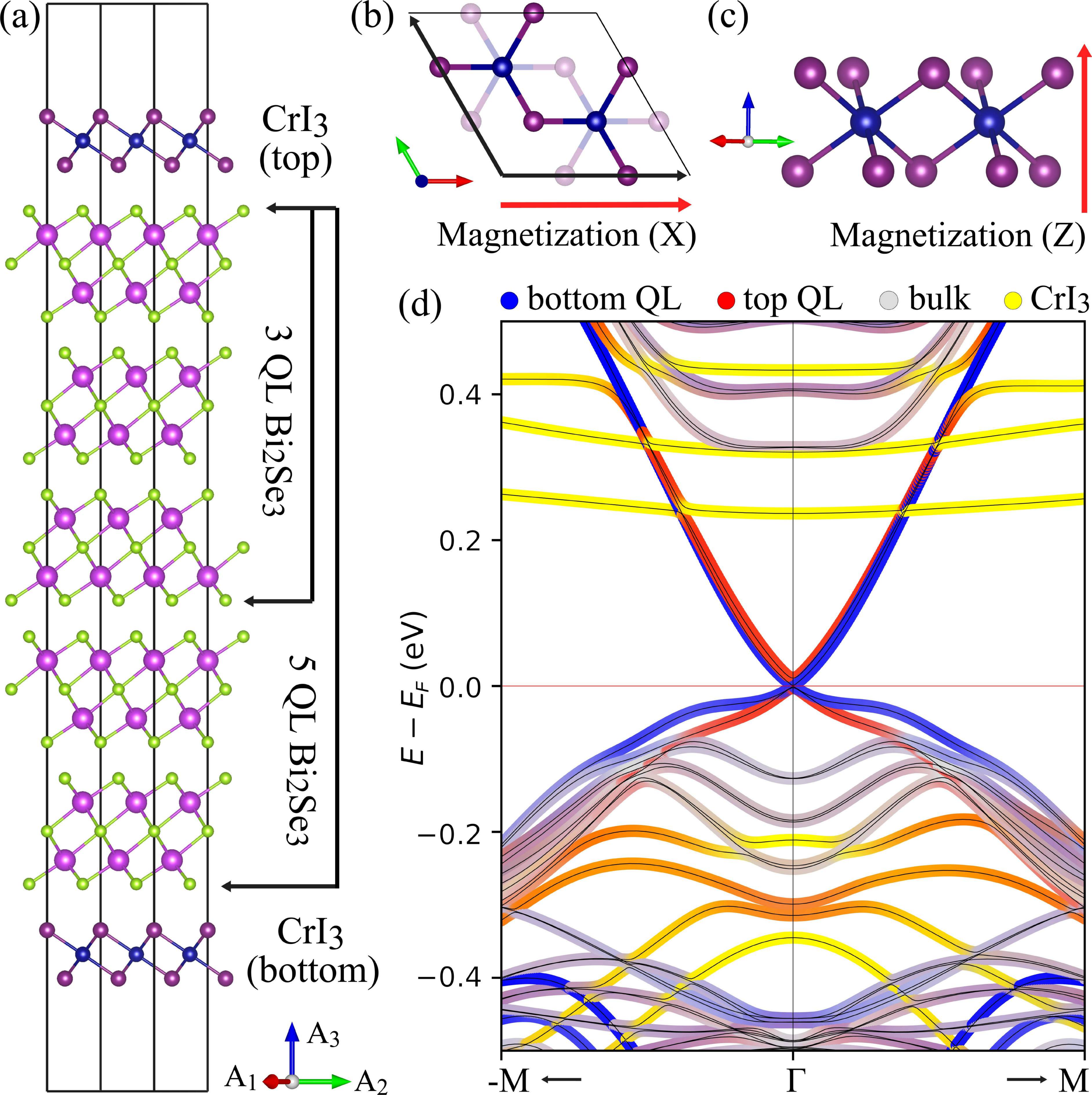}
    \caption{\label{fig:Lattice} (a) Side view of the CrI$_3$/Bi$_2$Se$_3$ interface. CrI$_3$ layer, indicating the in-plane (b) and out-of-plane (c) magnetization directions adopted in our calculations. (d) Electronic structure emphasizing the absence of overlap between CrI$_3$ and Bi$_2$Se$_3$ states in low-energy states (close to the Fermi energy $E_F$). The mixing of the color pattern indicates states with contributions coming from different regions of the material.}
\end{figure}

The CrI$_3$/Bi$_2$Se$_3$ interface is constructed by considering the CrI$_3$ layer on the surface of the Bi$_2$Se$_3$ with quintuple layers (QLs) stacking, as illustrated in Fig.~\ref{fig:Lattice}(a). To achieve a lattice match between the two materials, we adopt a supercell of $(\sqrt{3}\times\sqrt{3})$-Bi$_2$Se$_3$, with the CrI$_3$ layer slightly strained ($\sim 2.17\%$). It is worth pointing out that for such strain value, the CrI$_3$ system retains its ferromagnetic and insulating character. Among the possible alignments between the Bi$_2$Se$_3$ and CrI$_3$ layers, we use the lowest energy configuration, namely, that with the Cr atom positioned atop the Se atom \cite{CrI3-5}. The optimized separation distance between the two materials is $3.07$\,{\AA}.

\subsection{Density functional theory}
\label{sec:density-functional-theory}

The first principles calculations are performed using the density functional theory (DFT) \cite{dft1, dft2} within the generalized gradient approximation (GGA) for the exchange and correlation functional, employing the Perdew-Burke-Ernzerhof (PBE) parametrization \cite{PBE}. A fully relativistic $j$-dependent pseudopotential, within the projector augmented wave method (PAW) \cite{PRBblochl1994, Corso2010}, was used in the noncollinear spin-DFT formalism self-consistently. We use the Vienna {\it ab initio} Simulation Package (VASP) \cite{vasp1, vasp2}, with plane wave basis set with a cut-off energy of 400 eV. The Brillouin zone is sampled using a number of $k$-points such that the total energy converges within the meV scale. The vacuum space is set to $15$\,{\AA} and atomic structures are optimized requiring that the force on each atom to be less than $0.01$\,eV/{\AA}.

To accurately describe the strong relativistic effect produced by the Bi and I atoms, as well as the weak van der Waals interaction between the Bi$_2$Se$_3$ and CrI$_3$ layers, we include the SOC term \cite{PRBsteiner2016} and the nonlocal vdW functional (optB86b-vdW) \cite{VDW-optB86b}.  Finally, the strong exchange effect due to $d$-orbitals of the Cr atoms is accounted for by introducing an on-site Coulomb $U_{\rm eff} = 2.1$\,eV interaction within the L(S)DA+U approximation \cite{LSDA+U}.

The 2D spin-orbital textures are computed on a $21\times21$ $k$-points grid over the $k_xk_y$ plane in the BZ. More specifically, we study the 2D spin-orbital textures over wide and small reciprocal space regions around the $\Gamma$-point, namely, $-0.1 (2\pi/a)\le k_x, k_y \le 0.1(2\pi/a)$ and $-0.01 (2\pi/a)\le k_x, k_y \le 0.01(2\pi/a)$. The analysis of the DFT results, including band structure plots, surface/layer contribution projections, and spin textures along 2D Fermi surfaces, has been performed using the VASProcar post-processing code \cite{vasprocar}.

\subsection{Effective Hamiltonian}
\label{sec:effective-model}

Bi$_2$Se$_3$ is a 3D topological insulator with electronic surface states characterized by a Dirac-like dispersion and chiral spin textures, as shown by our DFT results (see, for instance, Figs.~\ref{fig:Lattice} and \ref{fig:spin_orbital}). 
The surface states are nicely described by the effective Hamiltonian model
\begin{equation}
\label{eq:H0}
 \hat{H}_0 = \hbar v_{\rm F} \left( {\bf k} \times \bm{\sigma} \right)\cdot \hat{\bf z} \otimes \tau_z   \,
\end{equation}
where ${\bm \sigma} = (\sigma_x, \sigma_y, \sigma_z)$ are Pauli matrices operating on the electron spin degree of freedom (up/down), whereas ${\bm \tau} = (\tau_x, \tau_y, \tau_z)$ act on the surface degree of freedom (top/bottom). For clarity, the $2 \times 2$ identity matrix is represented by $\sigma_0$ and $\tau_0$.

For thin slabs (3QLs) an inter-surface coupling opens an energy gap at the vertex of the Dirac cone. Such effect is captured by adding an inter-layer coupling term to the effective model, Eq.~\eqref{eq:H0}, namely
\begin{equation}
    \hat{H}_D = \delta \,\sigma_0 \otimes \tau_x \,,
\end{equation}
where $\delta$ stands for the inter-layer hybridization matrix element. 

At the interface, the magnetic moments of the CrI$_3$ couple to electronic spin of the topological surface states, changing both their energy dispersion as well as their spin texture. Figure~\ref{fig:Lattice}(d) shows the energy bands of a system formed by CrI$_3$ on the surface of Bi$_2$Se$_3$ 5QL, with out-of-plane magnetization. Despite the magnetic moment been located at the Cr atoms, the Bi$_2$Se$_3$ surface states effectively feel an almost uniform magnetic field. Therefore, we can add an uniform magnetic field term $H_M$ to the effective model Hamiltonian
\begin{equation}
    \hat{H}_M = 
    \sum_{i=t,b} \bm{m}_{i} \cdot \bm{\sigma} \otimes \tau_i
\end{equation}
here $\bm{m}_i = \hbar \mu_b \bm{B}_i/2$, where $\bm{B}_i$ is the effective external magnetic field of the CrI$_3$ in the top ($i=t$) and botton ($i=b$) interfaces, and $\mu_b$ the Bohr magneton.. Note also that $\tau_{t/b} = \left( \tau_z \pm \tau_0 \right)/2$ project the surface subspace, and therefore $\hat{H}_M$ can describe a single or both surfaces interacting with a uniform magnetic field. Given the possible asymmetric configuration between top/botton surfaces, we also add an inversion symmetry breaking term
\begin{equation}
    \hat{V} = U \sigma_0 \otimes \tau_z,
\end{equation}
with $U=eV$, $e$ the electron charge and $V$ the potential difference between surfaces. Here, the inversion symmetry breaking term represents a dipole moment originated by the presence of CrI$_3$.  Similar effect can also be induced by an external electric field due to a gate.

We can write the low energy Hamiltonian in the basis $\{ |t,\uparrow\rangle, |t,\downarrow\rangle, |b,\uparrow\rangle, |b,\downarrow\rangle \}$, where $t\; (b)$ stand for the top (bottom) surface and $\uparrow (\downarrow)$ for spin up (down) in the $\sigma_z$ basis, as 
{\footnotesize
\begin{equation*}
      H = \begin{pmatrix}
      U + m_{t,z}    &  m_{t,x} + \hbar v_F \tilde{k}^*  &   \delta       &   0            \\ 
      m_{t,x} + \hbar v_F \tilde{k}  &  U - m_{t,z}    &   0            &   \delta       \\
      \delta       &  0            &  -U + m_{b,z}   &  m_{b,x} - \hbar v_F \tilde{k}^*  \\
      0            &  \delta      &  m_{b,x} - \hbar v_F \tilde{k}  &  -U - m_{b,z}
     \end{pmatrix},
\end{equation*} }
where we define $\tilde{k} = k_y + ik_x = k e^{i \theta}$ with $\theta$ with respect to the $k_y$ direction. 

\begin{figure}
    \includegraphics[width=0.95\columnwidth]{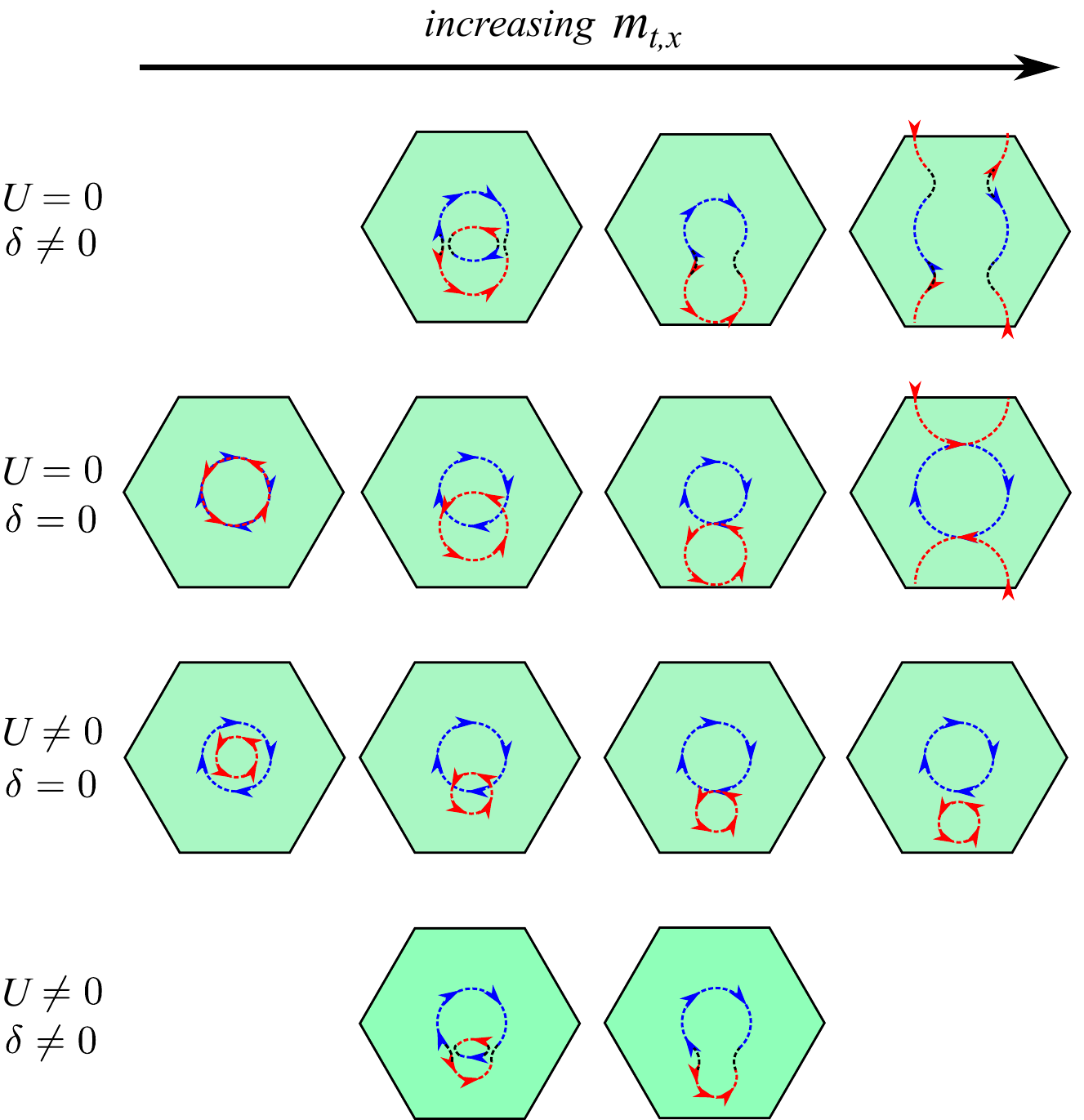}
    \caption{\label{fig:model} Family of spin textures and 2D Fermi surfaces generated by the effective Hamiltonian, by modulating thedifferent degrees of freedom of the TI/MI interface.}
\end{figure}

Figure~\ref{fig:model} shows a plethora of spin textures in the 2D Fermi surfaces obtained from the study of the effective model, for different possible parameters. For the CrI$_3$/Bi$_2$Se$_3$ interfaces, several of the effective Hamiltonian parameters are extracted from DFT calculations and are, from now on, fixed. Particularly for the CrI$_3$/Bi$_2$Se$_3$ interfaces we have extracted from the density functional theory calculations each parameter for the effective Hamiltonian. For instance, $\hbar v_F = 1.5$\,eV for all cases; $\delta = 7.5$\,meV and $0.1$\,meV for the 3QL and 5QL slabs, respectively;  $U=19.5$\,meV and $4.9$\,meV for single surface interfaces 3QL and 5QL slabs, respectively, and zero for both surfaces interface, {\it i.e.} CrI$_3$/Bi$_2$Se$_3$/CrI$_3$; the magnetization term $m_{\beta,i} = 6.4$\,meV for $\beta=t,\,b$  and $i=x,\,z$.

\subsection{Scattering Probability}
\label{sec:sscattering_probability}

In order to estimate the robustness of the different spin texture configurations against disorder, we study the scattering probability from a given initial state $|n_1,{\bf k}_1 \rangle$ to a final state $|n_2,{\bf k}_2 \rangle$, where $n_1,n_2$ stand for band indices and ${\bf k }_1,\,{\bf k}_2$ for the crystal momenta. For a given impurity potential $\hat{H}'$, we estimate for the transition probability between initial and final states using the Fermi golden rule
\begin{equation} 
    \Gamma_{i \rightarrow f} = \frac{2\pi}{\hbar} |\langle n_2,{\bf k}_2 | \hat{H}' | n_1,{\bf k}_1 \rangle|^2 \rho(E_{n_2,{\bf k}_2}),
\end{equation}
where $\rho(E)$ is the density of states at the energy $E$.

Since we are interested in general properties of elastic scattering processes further simplifications is in order. 
To avoid addressing specifics of the electron-impurity cross sections, we consider disorder processes that does not hybridize different surface states and non-magnetic impurities, that do not flip spin.
Further, we address the long range disorder limit, for which the scattering lengths $a_s$ are much larger than the electron wave length $\lambda$, namely, $a_s \gg \lambda$. 
Hence, we can approximate the transition matrix squared by a product of a function that depends on some average impurity potential times a 4-dimensional spinor (spin+surface) projection and write
\begin{equation} 
     |\langle n_2,{\bf k}_2 | \hat{H}' | n_1,{\bf k}_1 \rangle|^2 \propto |\langle n_2,{\bf k}_2 | n_1,{\bf k}_1 \rangle|^2.
\end{equation}
Consequently, $\Gamma_{i \rightarrow f} \propto |\langle n_2,{\bf k}_2 | n_1,{\bf k}_1 \rangle|^2$.
These approximations allow us to study the impurity independent term and to infer some anisotropic features of disorder scattering processes.

\section{Results and discussion}
\label{sec:Results-discussion}
    
Here we present a comprehensive {\it ab initio} characterization of the electronic structure and spin-orbital texture behavior of CrI$_3$/Bi$_2$Se$_3$ interfaces. In addition, we interpret the underlying physics in term of the effective Hamiltonian and scattering probabilities introduced above.

Our calculations explore different system settings and configurations, namely, ($i$) finite-size effects caused by a TI slab width, ($ii$) the magnetization direction of the FM, and ($iii$) the presence of CrI$_3$ in the slab interface (single or both surfaces). The interplay of the effects corresponding to the later settings give rise to a rich variety of emergent spin-orbital texture configurations. 

\subsection{Low-Energy Dispersion and 2D Fermi Surfaces}
\label{sec:energy_dispersion}

\begin{figure}[h]
    \FloatBarrier\includegraphics[width=1.0\columnwidth]{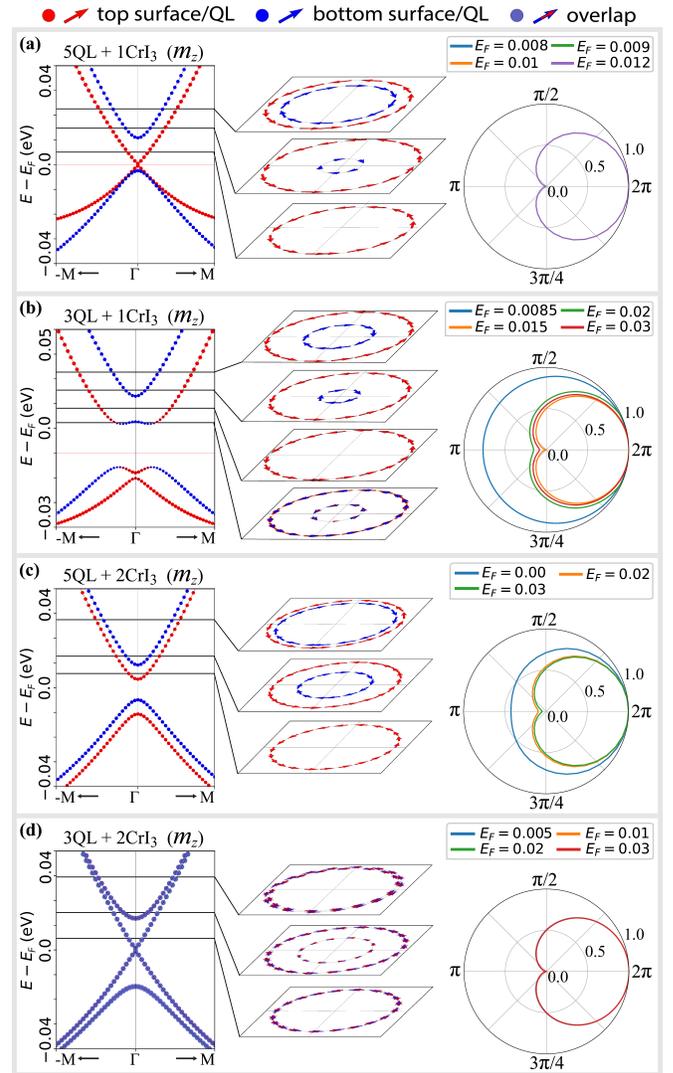}
    \caption{\label{fig:FS-spin-DFT_b} Out-of-plane magnetization. Projection of the top and bottom surfaces contribution to the low energy dispersion. Spin textures over the 2D Fermi surfaces, where the size of the arrows and the colors represent the contribution surfaces to the spin vector $(S_x,\,S_y)$ in the $(k_x,\,k_y)$ plane. Scattering probability, for different Fermi energies (in eV).}   
\end{figure}

The control of finite-size effects is achieved by varying the thickness of the Bi$_2$Se$_3$ stacking. We show that, while there is a significant surface states hybridization in 3 QLs systems, the coupling is practically absent in 5 QLs. This is inferred by observing that for thinner slabs (3 QLs), the topological surface states from opposing surfaces are coupled, opening an energy gap at the crossing of states with the same spin. This overlap is suppressed in 5QLs slabs ($\delta \sim 0$). For instance, comparing Fig.~\ref{fig:FS-spin-DFT_b}(a) with Fig.~\ref{fig:FS-spin-DFT_b}(b), corresponding to 5QL and 3QL systems, respectively, (keeping other degrees of freedom fix) we see a gap opening and a superposition between top (red) and bottom (blue) surface states. In those systems, there is a CrI$_3$/Bi$_2$Se$_3$ interface only at one of the Bi$_2$Se$_3$ slab surfaces, with the CrI$_3$ magnetization normal to the interface, namely, $n$QL+1CrI$_3$($m_z$) with $n =$ 3 and 5.

Regarding the magnetization direction of CrI$_3$, we analyzed the magnetizations in-plane ($m_x$) and out-of-plane ($m_z$). The main effect of $m_z$ magnetization is to induce an energy gap in the surface states adjacent to the CrI$_3$ layer, due to the introduction of coupling between states of the same surface and opposite spins (terms $m_{t/b,z}$  in the Hamiltonian). Additionally, there is a Zeeman-like energy shift of the Dirac states, lifting the degeneracy of the surface states, as shown in Fig.~\ref{fig:FS-spin-DFT_b}.

\begin{figure}[h]
    \FloatBarrier\includegraphics[width=1.0\columnwidth]{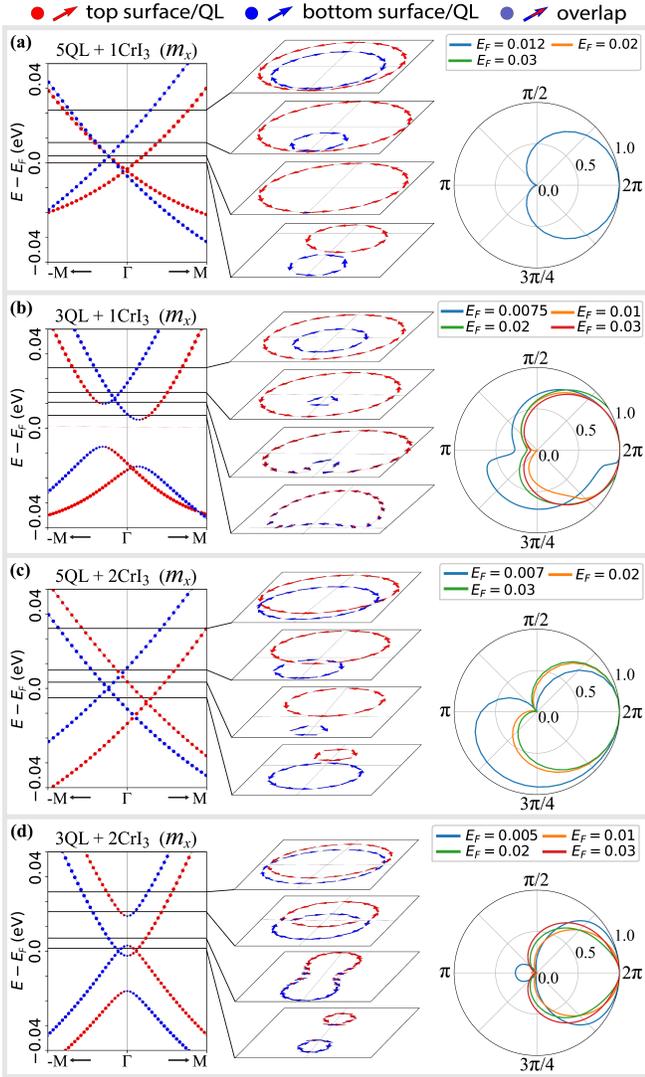}
    \caption{\label{fig:FS-spin-DFT_a} In-plane magnetization. Projection of the top and bottom surfaces contribution to the low energy dispersion. Spin textures over the 2D Fermi surfaces, where the size of the arrows and the colors represent the contribution surfaces to the spin vector $(S_x,\,S_y)$ in the $(k_x,\,k_y)$ plane. Scattering probability, for different Fermi energies (in eV).}   
\end{figure}

In turn, the in-plane $m_x$ magnetization preserves the crossing of the Dirac cones. However, an efficient switching occurs between the magnetic moment direction and the adjacent surface states, resulting in a momentum shift of the states perpendicular to the magnetization direction, whose direction is associated with helicity of the spin texture, leading to the breaking of Fermi contours concentricity, as shown in Fig.~\ref{fig:FS-spin-DFT_a}. Hence, by modulating the intensity and direction of magnetization, allow for the control of the direction and shape of both the energy dispersion and the spin component orientation of the topological states of Bi$_2$Se$_3$, a feature which is potentially advantageous for spintronic applications. Conversely, the reciprocal effect of this coupling can enable electrical control of the magnetization direction of CrI$_3$ when a charge current is introduced in Bi$_2$Se$_3$, given the well stablished presence of the spin-orbit torque (SOT) effect mediated by REE in TI/FM interfaces with in-plane magnetization \cite{SOT, STT-1}, with potential application for the development of magnetic memories.

\subsection{Spin-Orbital Textures}
\label{sec:spin_orbital_textures}

\begin{figure}[h!]
    \includegraphics[width=0.9\columnwidth]{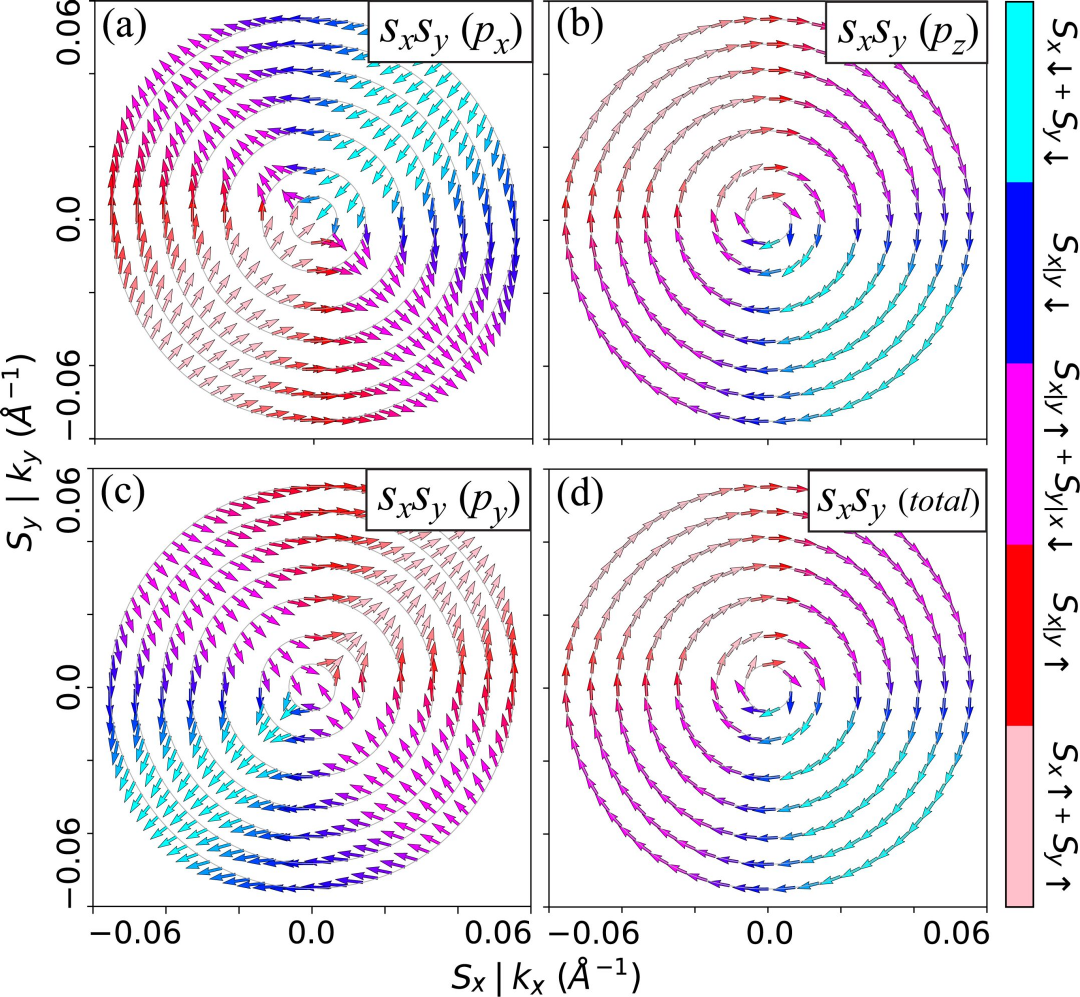}
    \caption{\label{fig:spin_orbital} Spin-orbital texture (near the $\Gamma$ point) corresponding to the $p_x$, $p_y$, and $p_z$ orbitals of the surface states of a 5 QL stacking of Bi$_2$Se$_3$.}
\end{figure}

The surface states of Bi$_2$Se$_3$ slabs show a very strong contribution from the $p_z$ orbital, which is responsible for their helical spin texture, with a spin-momentum locking characteristic of topological states. Meanwhile, the in-plane $p_x$ and $p_y$ orbitals induce more complex and opposing spin textures, which cancel each other out, as seen in Fig.~\ref{fig:spin_orbital} for the upper Dirac cone of a 5QL stacking of Bi$_2$Se$_3$. Interestingly, several studies have reported \cite{zhang2013spin, zhu2013layer, cao2013mapping, xie2014orbital, waugh2016minimal} that in Bi$_2$Se$_3$, in addition to the usual spin-momentum locking, the spin texture is tied to the orbital texture, giving rise to the so-called spin-orbital texture, with the $p_z$ orbital perpendicular to the layer's plane, while the $p_x$ and $p_y$ orbitals respectively form a radial and tangential pattern for the upper Dirac cone, and vice-versa for the lower cone.

For each of the analyzed CrI$_3$/Bi$_2$Se$_3$ interface configurations, we have examined the behavior of the spin-orbital textures arising from the $p_x$, $p_y$, and $p_z$ orbitals. By comparison with the texture pattern originating from the stacking of pristine Bi$_2$Se$_3$ (see Fig. \ref{fig:spin_orbital}), we have verified that the main effects consist of the relative shifts in energy and momentum of the textures derived from opposite surfaces, as well as a blending of the textures patterns due to overlapping states. However, the predominant features of the spin texture from the $p_z$ orbital, and the mutual cancellation of textures related to the $p_x$ and $p_y$ orbitals, remain unchanged, indicating that the magnetization originating from the Cr atom does not substantially modify the orbital distribution of the topological states of Bi$_2$Se$_3$. Due to the fact that the spin-orbital texture associated with the $p_z$ full is equivalent to the complete spin texture of the topological state, and therefore relevant to scattering processes. 

Regarding the $p_z$ spin-orbital texture, all the configurations we analyzed preserve the spin-momentum locking at the 2D Fermi surface, except for the 3QL+2CrI$_3$($m_x$) configuration [see the Fig.\ref{fig:FS-spin-DFT_a}(d)]. For this specific case, we observe that the spin configuration near the $\Gamma$ point becomes uniform in space and independent of momentum, similar to a persistent spin helix (PSH). In addition to obtaining the PSH state, we verify that the resulting spin texture aligns perpendicularly to the magnetization direction. Thus, it can be controlled through the tuning of the in-plane magnetization direction. This result aligns with one of the challenges in spintronics, which is obtaining a material where PSH is an intrinsic, controllable, and robust property of the system.

Another unusual spin texture occurs for a similar configuration, now with out-of-plane magnetization [see the Fig.~\ref{fig:FS-spin-DFT_b}(d)]. In the 3QL+2CrI$_3$($m_z$) configuration, the combination of $m_z$ magnetization on both surfaces with the hybridization effect leads each state to be distributed throughout the Bi$_2$Se$_3$ stacking and exhibit a null net spin texture. However, when projecting the spin components onto each surface, we observe spin textures with opposite chiralities, similar to those of the non-hybridized surface state, but here, each state has equal contributions from both surfaces.

\begin{figure}
    \FloatBarrier\includegraphics[width=0.992\columnwidth]{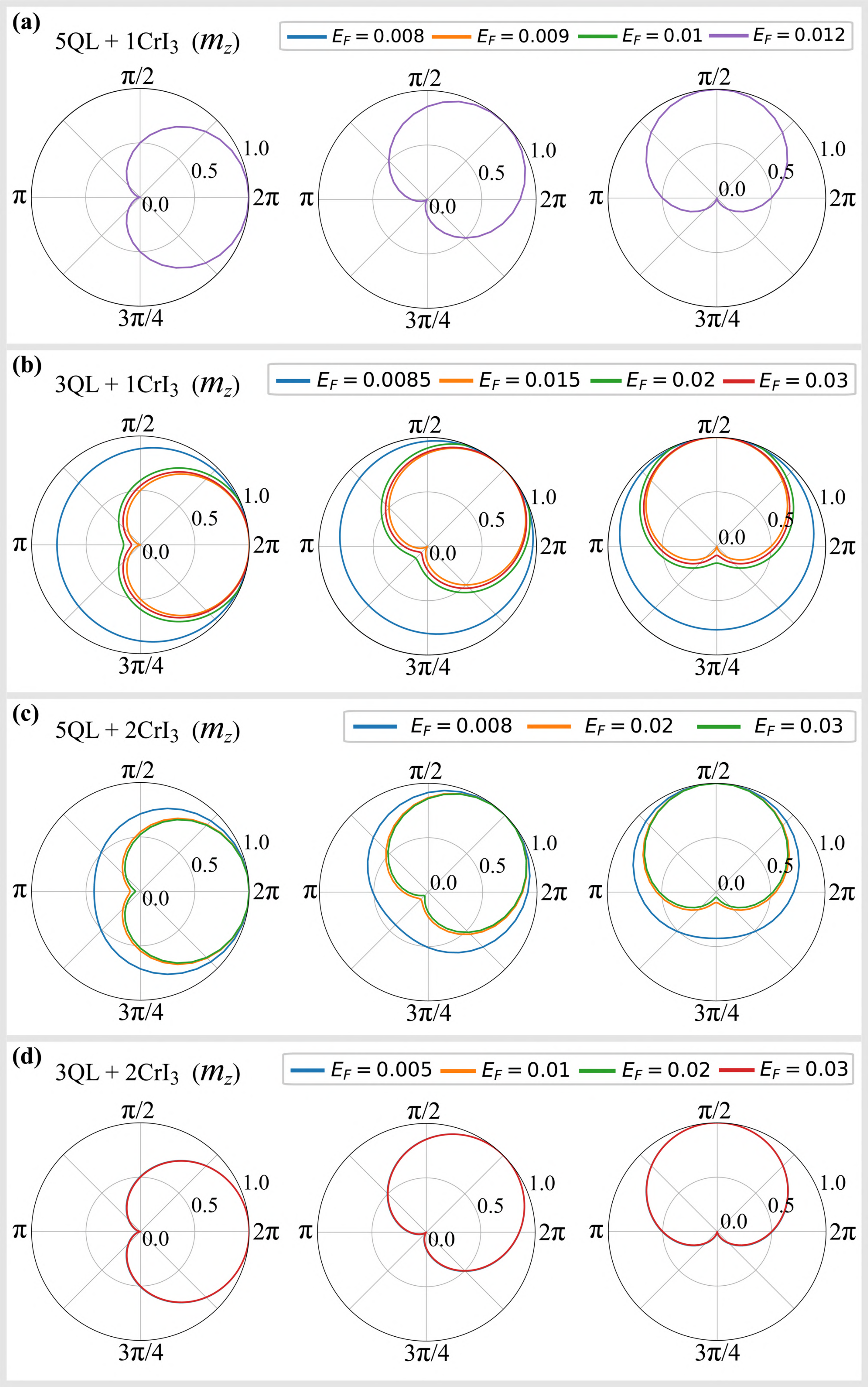}
    \caption{\label{fig:scattering_b} Out-of-plane magnetization: Scattering probabilities as a function of $E_F$ and propagation direction of the state along the BZ (each figure within the same frame refers to a specific position at moment $\vec{k}_1$ of the initial state, respectively, forming angles of 0$^{\circ}$, 45$^{\circ}$ and 90$^{\circ}$ with respect to the $k_x$ axis).}   
\end{figure}

\begin{figure}
    \FloatBarrier\includegraphics[width=\columnwidth]{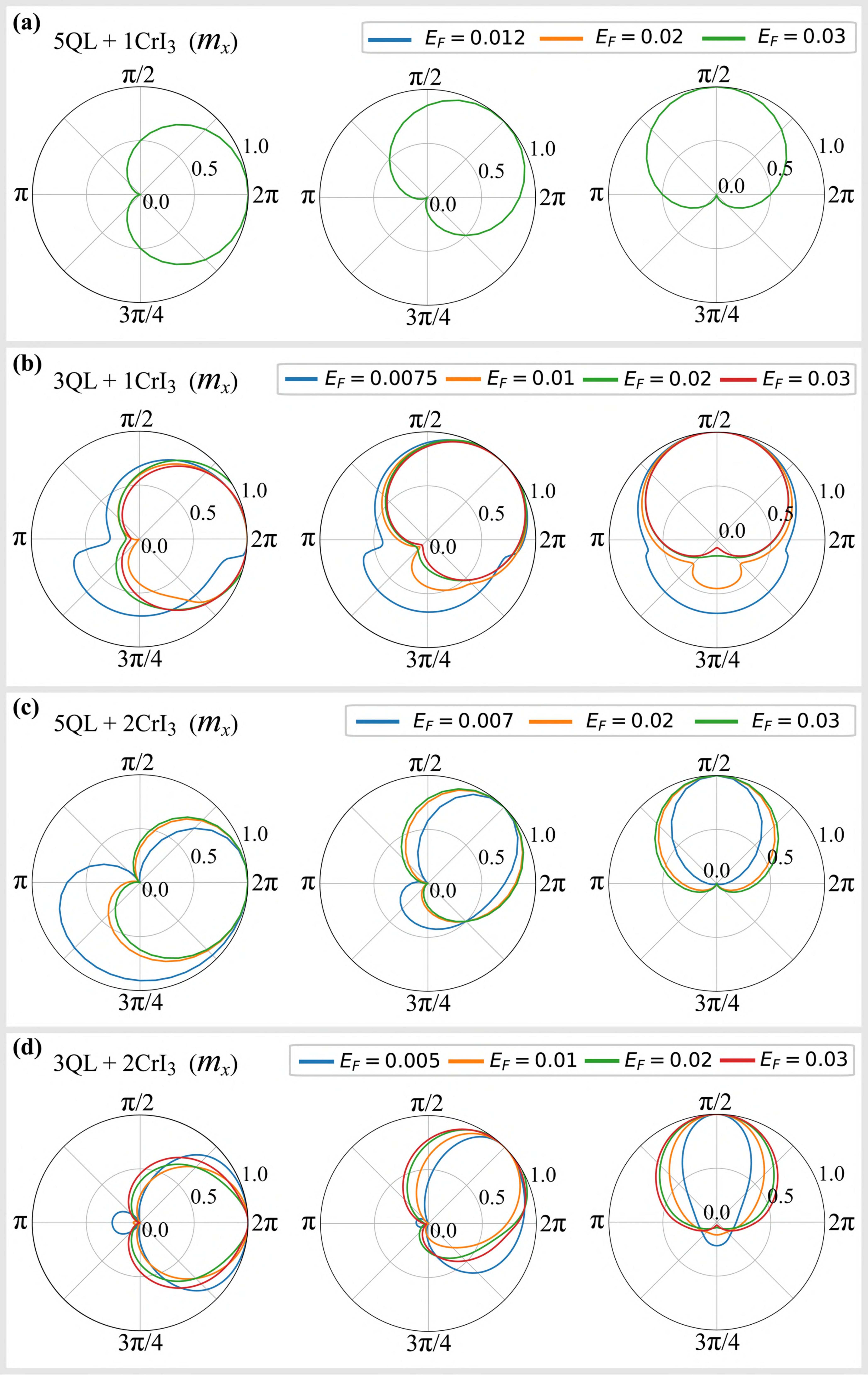}
    \caption{\label{fig:scattering_a} In-plane magnetization: Scattering probabilities as a function of $E_F$ and deflection angle of the state along the BZ. Each figure within the same frame refers to a different orientation of the initial state ${\bf k}_1$ namely, 0$^{\circ}$, 45$^{\circ}$ and 90$^{\circ}$ with respect to the $k_x$ axis).}     
\end{figure}

\subsection{Anisotropic scattering}

Here, we analyze the scattering profile by setting the incomming state $|n_1,{\bf k}_1 \rangle$ as a top surface state (colored in red in Figs.~\ref{fig:FS-spin-DFT_b} and \ref{fig:FS-spin-DFT_a}). We summarized our results in Figs.~\ref{fig:scattering_b} and ~\ref{fig:scattering_a}, depicting the scattering profile for out-of-plane and in-plane CrI$_3$ magnetizations, respectively. The most notable features are an isotropic behavior of the scattering for the out-of-plane magnetization ($m_z$) in Fig.~\ref{fig:scattering_b} and an anisotropic behavior for the in-plane magnetization ($m_x$) in Fig.~\ref{fig:scattering_a}.

For interfaces with in-plane magnetization, we find that the combination of the $k$-moment shift of the Dirac cones on the surface covered by CrI$_3$, with the overlapping of the wave functions between opposite surface states, promotes a strong anisotropy of the scattering profile as a function of the propagation direction of the initial state, which intensifies in the low energy limit (see the Figs.~\ref{fig:FS-spin-DFT_a} and \ref{fig:scattering_a}). However, the 5QL+1CrI$_3$($m_x$) configuration [Fig.~\ref{fig:FS-spin-DFT_a}(a)] does not present anisotropy, given the non displaced Dirac cone at $k$-moment (CrI$_3$ free surface). For this specific case, the inter-surface overlap term ($\delta$) does not affect the scattering, as (i) the intersection between states (above the Fermi level) only occurs at a single $k$-point along the entire BZ and (ii) is smaller for 5QL system.

Back scattering is expected to be favored in 3QL stacking due to inter-surface hybridization caused by finite size effects. This is in line with the results shown in Figs.~\ref{fig:scattering_b}(b) and \ref{fig:scattering_a}(b) for $E_F = 0.012$\,eV. At larger values of $E_F$, backscattering is suppressed. Interestingly, the 3QL+2CrI$_3$($m_z$) configuration [Fig.~\ref{fig:FS-spin-DFT_b}(d)], deviates from tis behavior and show an unusual and robust protection against backscattering for any analyzed energy value. Here a resonance occurs between the interlayer coupling term $\delta$ and the magnetization terms $m_{t/b,z}$, from DFT calculations we verify $\delta \approx m_{t/b,z}$. This resonance introduces an additional symmetry in the system: although time-reversal (TR) is broken a combination of TR and inversion is still preserved ([$\mathcal{I} \mathcal{T}$]) leading to the backscattering protection. On the other hand, in systems with 5QL backscattering protection is due to the suppression of overlap between surface states with opposite spin chiralities, as seen in the Figs.~\ref{fig:FS-spin-DFT_b} and \ref{fig:FS-spin-DFT_a}. However, in the 5QL+2CrI$_3$($m_z$) configuration [Fig.~\ref{fig:FS-spin-DFT_b}(c)] in the region near the $\Gamma$-point, the states acquire a finite $S_z$ component with the same direction, which in addition to the small (but finite) overlap between the wave functions, results in the breakdown of protection against backscattering.

Finally, we emphasize that the effects mentioned above are more prominent in the low energy limit, as the state is shifted to higher energies, the effects of the coupling terms cease to be significant, with the behavior of the pristine topological state (mainly protection against backscattering), being gradually re-established.

\section{Conclusions}
\label{sec:conclusions}

In conclusion, we have demonstrate that the surface states of a 3D topological insulator can give origin of a rich family of unconventional spin-orbital texture configurations on the 2D Fermi surfaces when in contact with a 2D ferromagnetic insulator. Further, we show that such van der Waals interface provides simple handler to control over the spin degree of freedom. By mapping each experimentally achievable texture configuration and estimating the corresponding scattering rates, we observe that the scattering of states due to the breaking of time-reversal symmetry is significant only at low energy scales ($\sim 10$\,meV). The expected robustness of topological states is reestablished at higher energy values, even in the presence of magnetization.

Additionally, we note that magnetization in the plane of the ferromagnetic layers induces anisotropy in transport phenomena and efficient switching between the directions of the magnetic moment and spin orientation. Furthermore, we find that properties of interest in spintronics can be obtained for specific interface configurations when finite-sized effects are present. These include robust protection against backscattering in the presence of magnetization [3QL+2CrI$_3$ ($m_z$)] and a unidirectional spin configuration independent of momentum near the $\Gamma$ point [3QL+2Cr$_3$ ($m_x$)].

We believe that these results can provide an efficient guide for manipulating spin-polarized states in topological systems, aiming to enhance control over emerging surface phenomena. This includes improving spintronic mechanisms directly related to spin texture configuration, such as charge/spin interconversion, spin relaxation time suppression, and spin torque transfer, as well as aiding in the modulation of transport properties and the development of devices with lower energy loss.

\begin{acknowledgments}

The authors acknowledge financial support from the Brazilian agencies FAPESP (grants 22/08478-6, 19/20857-0, and 17/02317-2), CNPq (INCT - Materials Informatics and INCT - Nanocarbono), FAPERJ, and LNCC - Laboratório Nacional de Computação Científica for computer time (projects ScafMat2 and EMT2D).

\newpage

\end{acknowledgments}


\bibliography{bib}

\end{document}